\documentclass[a4paper,12pt]{article}
\usepackage{amsmath}
\usepackage{amssymb}
\usepackage{amsfonts}
\usepackage{mathrsfs}
\usepackage{cite}
\usepackage[font=small,labelfont=bf]{caption}
\usepackage{graphicx}

\def\beq{\begin{equation}}
\def\eeq{\end{equation}}
\def\bea{\begin{eqnarray}}
\def\eea{\end{eqnarray}}

\def\<{\left\langle}
\def\>{\right\rangle}

\begin{document}

\bibliographystyle{OurBibTeX}

\begin{titlepage}

\vspace*{-15mm}

\begin{flushright}
SHEP-10-15\\
\end{flushright}

\vspace*{10mm}

\begin{center}
{
\sffamily
\Huge
SUSY ${SU(5)}$ with singlet plus adjoint matter \\
\vspace{0.1in}
and ${{A_4}}$ family symmetry
}
\\[12mm]
Iain~K.~Cooper\footnote{E-mail: \texttt{ikc1g08@soton.ac.uk}},
Stephen~F.~King\footnote{E-mail: \texttt{sfk@hep.phys.soton.ac.uk}}~and~
Christoph~Luhn\footnote{E-mail: \texttt{christoph.luhn@soton.ac.uk}}
\\[7mm]
{\small\it
School of Physics and Astronomy,
University of Southampton,\\
Southampton, SO17 1BJ, U.K.
}
\end{center}
\vspace*{1.00cm}

\begin{abstract}

\noindent
We propose a supersymmetric (SUSY) $SU(5)$ Grand Unified Theory (GUT)
including a single right-handed neutrino singlet and an adjoint matter
representation below the GUT scale and extend this model to include an $A_4$
family symmetry and a gauged anomaly-free Abelian group. In our approach
hierarchical neutrino masses result from a combined type~I and type~III seesaw
mechanism, and the $A_4$ symmetry leads to tri-bimaximal mixing which arises
indirectly. The mixing between the single right-handed neutrino and the
matter in the adjoint is forbidden by excluding an adjoint Higgs, leading to a
diagonal heavy Majorana sector as required by constrained sequential
dominance. The model also reproduces a realistic description of quark and
charged lepton masses and quark mixings, including the Georgi-Jarlskog
relations and the leptonic mixing sum rules $s=r\cos \delta$ and $a=-r^2/4$
with $r=\theta_C/3$. 
\end{abstract}

\end{titlepage}
\newpage
\setcounter{footnote}{0}

\section{Introduction}

There has been much evidence to suggest that the Standard Model (SM) is not a
complete description of particle physics and needs extending. Arguably one of
the most important pieces of experimental evidence for physics beyond the SM
is the measurement of small but non-zero neutrino mass, leading to theories of
neutrino mass and mixing. In the seesaw mechanism, a natural explanation for
such tiny neutrino masses is provided by the exchange of a heavy particle
leading to Majorana neutrino masses suppressed by the large mass of the
exchanged particle. The heavy particle in the seesaw mechanism must be a
colour singlet but can be an electroweak singlet fermion with zero
hypercharge, an electroweak triplet Higgs scalar with two units of
hypercharge, or an electroweak triplet fermion with zero hypercharge,
corresponding to the type~I \cite{Minkowski:1977type1}, type~II \cite{type2},
or type~III \cite{Foot:1989type3} seesaw mechanisms, respectively. In this
letter we shall combine the seesaw mechanism of types~I and~III in a new way
to yield a hierarchical spectrum of neutrino masses.

However the seesaw mechanism is not by itself enough to account for the
discovery that, in contrast with the smallness of the quark mixing angles, two
out of the three leptonic mixing angles are large. This unexpected
observation calls for a deeper theoretical understanding of the physics
underlying the structure of fermion masses and mixings. It is well known that
solar and atmospheric data are consistent with a simple form of the leptonic
mixing matrix $U$, known as tri-bimaximal (TB) mixing \cite{Harrison:2002tbm}:
\beq
 U_{TB}=\begin{pmatrix}
         \sqrt{\frac{2}{3}} & \frac{1}{\sqrt{3}} & 0 \\
	 -\frac{1}{\sqrt{6}} & \frac{1}{\sqrt{3}} & -\frac{1}{\sqrt{2}} \\
	 -\frac{1}{\sqrt{6}} & \frac{1}{\sqrt{3}} & \frac{1}{\sqrt{2}}
        \end{pmatrix}
\label{eqn:tbm}.
\eeq
The simple form of this matrix can be interpreted as a clue that points
towards some underlying family symmetry $G_f$, related to particular
transformations which leave the mass matrix diagonalised by $U_{TB}$
invariant. There has been much recent work based on this idea that the
postulated TB symmetry can arise from a family symmetry
\cite{S3-L,Dn-L,A4-L,A4-L-Altarelli,S4-L,delta54-L,S3-LQ,Dn-LQ,Qn-LQ,A4-LQ,King:2007A4,doubleA4-LQ,S4-LQ,S4-LQsum,PSL-LQ,Z7Z3-LQ,delta27-LQ,delta27-LQ-Dterms,SO3-LQ,SU3-LQ,Reviews}. The
approaches taken in the literature may be separated into two distinct classes,
distinguished by the breaking of the family symmetry \cite{King:2009luhn2}:
\emph{direct} models, based on $A_4$, $S_4$, or larger groups containing these
as subgroups \cite{Lam,PSLtheory}, have part of the family symmetry preserved
at low energies and this forms some or all of the neutrino flavour symmetry;
\emph{indirect} models, usually based on $\Delta(3n^2)$ or $\Delta(6n^2)$
\cite{deltastheory}, have entirely broken family symmetries (in the neutrino
sector), and the neutrino flavour symmetry appears accidentally.

Finding an explanation for the distinctly different mixing patterns of leptons
as compared to quarks is even more important in the context of Grand Unified
Theories (GUTs) \cite{Langacker:1980js,Amsler:2008zzb}, where the fermionic
matter is unified at high energies into either a single representation, as in
$SO(10)$~\cite{Fritzsch:1974nn} or $E_6$~\cite{Gursey:1975ki}, 
or into two representations, as in
$SU(5)$~\cite{Georgi:1974GUT} or $SU(4)_{PS}\times SU(2)_L\times
SU(2)_R$~\cite{Pati:1973uk}. The minimal GUT~\cite{Georgi:1974GUT} is based on
the Lie group $SU(5)$, where one family of right-handed down quarks and
left-handed leptons are unified in a ${\bf \overline{5}}$ and the rest of the
family are in a ${\bf 10}$. Three copies of each of these 
representations then constitutes the full fermionic matter content of minimal
$SU(5)$. It is well known that gauge coupling unification fails in this
regime, however if promoted to a supersymmetric (SUSY) GUT~\cite{SUSYguts} with two Higgs multiplets, $H_{\bf 5}$ and~$H_{\bf
  \overline 5}$, then unification occurs at a scale of roughly
$2\times10^{16}$ GeV \cite{Langacker:1990GUT}. 

It is clear that neutrino masses are zero at the renormalisable level in
minimal (SUSY) $SU(5)$ as in the minimal SM. However, in both the SM and
$SU(5)$, as pointed out by Weinberg \cite{dim5}, one may invoke a
non-renormalisable dimension-5 operator at or above the GUT scale to generate
neutrino masses. Such an operator at the GUT scale may be sufficient to
describe the solar neutrino mass scale, but not the atmospheric neutrino mass
scale. In order for neutrinos to obtain mass consistent with atmospheric
mixing in a (SUSY) $SU(5)$ model, the seesaw mechanism is a very attractive
possibility, however this requires some extra matter or Higgs to be added
below the GUT scale. The choice of additional matter or Higgs is very {\it ad
  hoc} since the $SU(5)$ theory does not specify the nature of this extra
matter and only requires that it be anomaly-free. A popular choice is to add
three right-handed neutrinos which arise from singlet $SU(5)$ representations.
However the number of singlets is not predicted in $SU(5)$, and it is possible
to add just a single right-handed neutrino to describe the atmospheric mass
scale \cite{King:1998jw}. In order to describe both atmospheric and solar
neutrino masses with two large mixing angles using the type~I seesaw mechanism
two right-handed neutrinos are sufficient \cite{2RHN}. However, within $SU(5)$
GUTs, there are other possibilities.

It has been pointed out that, in (SUSY) $SU(5)$ GUTs, non-fundamental matter
multiplets have decompositions which include both fermion singlets and fermion
triplets suitable for the type~I and III seesaw mechanism, the smallest such
example being the adjoint $\bf 24$ representation
\cite{Ma:1998adjoint,Dorsner:2006fx,Perez:2007susy}. The decomposition of a
matter $\bf 24$ under the SM gauge group $SU(3)_c\times SU(2)_L\times U(1)_Y$
involves an $SU(2)_L$ singlet $\rho_0 = ({\bf{1}},{\bf{1}})_0$ as well as a
triplet $\rho_3 = ({\bf{1}},{\bf{3}})_0$, thus leading to a combination of a
type~I seesaw with a type~III seesaw \cite{Foot:1989type3}. However, assuming
the simplest Higgs sector, the $\rho_0$ and $\rho_3$ are constrained by
$SU(5)$ to give equal contributions to the neutrino mass matrix, up to an
overall constant, resulting in a rank one neutrino mass matrix and only one
non-zero neutrino mass. This problem may be addressed by allowing additional
couplings to a Higgs $\bf 45$ \cite{Perez:2007susy}, but here we shall
consider a different possibility.

\begin{figure*}
\centerline{
\mbox{\includegraphics[width=6in]{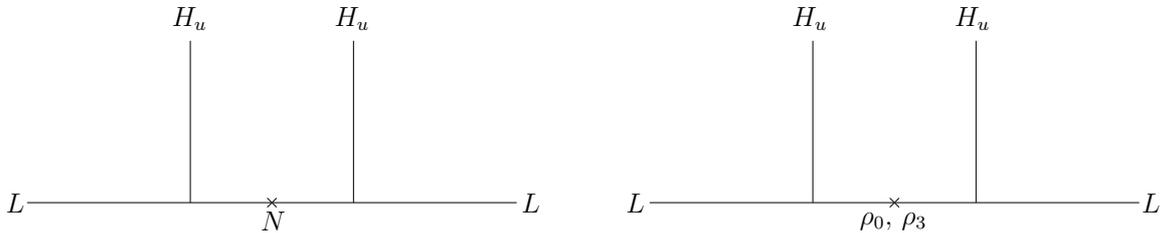}}
}
\caption{Schematic diagrams of the type~I (left) and combined type~I +
  type~III  (right) seesaw mechanisms present in the model. The seesaw
  messenger states are $N$ and the $\rho_0$,~$\rho_3$ components of $\psi_{\bf
  24}$. $L$ is the $SU(2)_L$ doublet contained in the ${\bf \overline 5}$ of
  $SU(5)$.}
\label{fig:seesaw}
\end{figure*}

In this letter we consider a SUSY $SU(5)$ GUT with one single
right-handed neutrino singlet superfield $N$ plus one adjoint matter
superfield $\psi_{\bf 24}$ below the GUT scale. The model combines a type~I
seesaw mechanism from the single right-handed neutrino $N$ below the GUT scale
\cite{King:1998jw} with a type~I plus type~III seesaw mechanism from the
$\rho_0$ and $\rho_3$ components contained in a single adjoint matter superfield
$\psi_{\bf 24}$ below the GUT scale \cite{Perez:2007susy}.
The seesaw mechanism in our model therefore results from three distinct diagrams
as shown in Fig.~\ref{fig:seesaw}. In order to describe TB mixing we also
include an $A_4$ family symmetry, plus an anomaly-free gauged $U(1)$ symmetry.
Instead of using an adjoint Higgs representation $H_{\bf 24}$ to spontaneously
break $SU(5)$ to the SM gauge group, we shall assume implicitly that 
the GUT group is broken by geometrical effects in
extra dimensions. However the theory here is formulated in four dimensions and 
we simply assume that it could subsequently be uplifted to
a higher dimensional setting (as in, for example, \cite{Albright:2002pt}).
The absence of $H_{\bf 24}$ is crucial in forbidding the 
mixing between the right-handed neutrino $N$ and $\psi_{\bf 24}$, 
leading to no mass mixing between
$N$ and $\rho_0$ and hence a diagonal heavy Majorana sector
as required by constrained sequential dominance (CSD)
\cite{King:2005CSD}. The flavon vacuum
alignments arise from the elegant $D$-term mechanism
\cite{Howl:2009ds}. The model also reproduces a realistic description of quark
and charged lepton masses and quark mixings, including the Georgi-Jarlskog
relations \cite{Georgi:1979Jarlskog}.

The remainder of this letter is organised as follows. In
Section~\ref{sec:model0} we introduce the SUSY $SU(5)$ model with singlet plus
adjoint matter below the GUT scale. Section~\ref{sec:model} describes the full
version of the model, including an $A_4$ family symmetry, a gauged $U(1)$
symmetry plus two discrete $Z_2$ symmetries, and lists the operators allowed
by these symmetries, resulting in the mass matrices for quarks, charged
leptons and neutrinos. We conclude in  Section~\ref{sec:conclusion}.



\section{SUSY ${\boldsymbol{SU(5)}}$ with singlet
 ${\boldsymbol \&}$ adjoint matter}
\label{sec:model0}
In this Section we consider a SUSY $SU(5)$ GUT with one single right-handed
neutrino arising from a singlet representation $N$ below the GUT scale plus
one extra adjoint matter representation $\psi_{\bf 24}$ with mass also below
the GUT scale. The matter contained in the $\psi_{\bf 24}$ is
degenerate thus avoiding problems with gauge coupling unification. The model
represents a new way to achieve a hierarchical neutrino mass spectrum arising
from a type~I plus type~III seesaw mechanism, as we now discuss.

The superpotential describing the neutrino sector takes the form
\begin{equation}
{W}=c_iF_i{\bf \psi_{24}}H_{\bf 5} + p_iF_i{N}H_{{\bf 5}} + \frac{1}{2}m_N NN + \frac{1}{2}m\,
\text{Tr}\,({\psi_{\bf 24}}^2),
\label{eqn:type3}
\end{equation}
with the $F_i$ denoting the three families of ${\bf \overline{5}}$s.
$N$ is a single right-handed Majorana neutrino superfield and ${\bf \psi_{24}}$ 
the additional adjoint matter superfield. 
The seesaw diagrams
illustrated in Fig.~\ref{fig:seesaw} then yield the light neutrino mass matrix,
\begin{equation}
\label{eqn:type3matrix}
M_{\nu}^{ij}={c_ic_j}v_u^2
\left(\frac{1}{4m_{\rho_{3}}}+\frac{3}{20 m_{\rho_{0}}}\right)
+  \frac{p_ip_j}{m_{N}}v_u^2 \, .
\end{equation}
Here $v_u$ is the vacuum expectation value (VEV) of the Minimal Supersymmetric
Standard Model (MSSM) Higgs field $H_u$ which corresponds to the $SU(2)_L$
doublet within the $SU(5)$ Higgs $H_{\bf 5}$. 
The numerical factors of the two
terms in parentheses are obtained by writing ${\bf   \psi_{24}}=
\tilde\rho_aT^a$, where $\tilde\rho_a$ are the 24 components of
the ${\bf \psi_{24}}$ and $T^a$ are the 24 appropriately normalised generators
of $SU(5)$ \cite{Langacker:1980js}. As can be seen from Eq.~(\ref{eqn:type3}),
the Majorana masses for the seesaw messengers $\rho_0$ and $\rho_3$ are
identical, i.e. $m_{\rho_0}=m_{\rho_3}=m$, while $N$ has an independent mass
$m_N$. Note that we have not introduced
an adjoint Higgs $H_{\bf 24}$ which would break the degeneracy of the components
in the $\psi_{\bf 24}$ and, more importantly, allow
a mixing term $N\psi_{\bf 24}H_{\bf 24}$ leading to a mass mixing between $N$ and $\rho_0$.
Note also that $c_i$ and $p_i$ are independent dimensionless coefficients
(where $i$ and $j$ are family indices); this
independence is crucial to obtaining a rank two mass matrix and thus two
non-zero neutrino masses.

As $c_i$ and $p_i$ are uncorrelated parameters, Eq.~(\ref{eqn:type3matrix})
does not in general conform to the TB structure of the neutrino mass
matrix. It is the aim of this letter to obtain TB neutrino mixing
as a consequence of a discrete family symmetry in this type of model. To this
end, in the next Section, we augment the adjoint SUSY $SU(5)$ model with the
tetrahedral family symmetry $A_4$.


\section{SUSY ${\boldsymbol{A_4 \times SU(5)}}$ with singlet ${\boldsymbol \&}$ adjoint matter}
In this Section we uplift the model in Eq.~(\ref{eqn:type3}) to include a
tetrahedral family symmetry. We work in the basis of \cite{King:2007A4} in
which two $A_4$ triplets $a=(a_1,a_2,a_3)^T$ and $b=(b_1,b_2,b_3)^T$ give a
singlet through the combination $a_1b_1+a_2b_2+a_3b_3$.
The basic idea is to unify the three families of ${\bf\overline{5}}$s into an
$A_4$ triplet~$F\sim {\bf 3}$, and in order for Eq.~\eqref{eqn:type3} to remain
invariant, to introduce extra $A_4$ triplets called flavons $\varphi_{i}$ to
break the $A_4$ symmetry and generate the Yukawa couplings.

Table~\ref{tab:indirect} shows the chiral superfields present in the model.
As mentioned above, the three ${\bf \overline{5}}$s of
$SU(5)$ are embedded in a triplet of $A_4$, while the three ${\bf 10}$s are
singlets. The ${\bf\psi_{24}}$ is an $A_4$ singlet as is the right-handed
neutrino $N$. We have fundamental Higgs fields $H_{\bf 5}$ and $H_{{\bf
    \overline{5}}}$; introducing another Higgs in the ${\bf \overline{45}}$
representation, $H_{{\bf \overline{45}}}$, enables the implementation of the
Georgi-Jarlskog mechanism \cite{Georgi:1979Jarlskog} to obtain the well known
GUT scale mass relations $m_e \sim \frac{m_d}{3}$, $m_{\mu}\sim 3m_{s}$ and
$m_{\tau}\sim m_b$. These give phenomenologically successful predictions of
down quark and charged lepton masses when evolved down to the electroweak
scale.

\begin{table}
	\centering
		\begin{tabular}{|c||c|c|c|c|c|c||c|c|c|c|c|c|c|c|c|c||c|}
			\hline
				Field & ${\bf \psi_{24}}$ & $N$ & $F$ & $T_1$ & $T_2$ & $T_3$ & $H_{\bf 5}$ & $H_{{\bf \overline{5}}}$ & $H_{{\bf \overline{45}}}$ & $\varphi_{123}$ & $\varphi_{23}$ & $\varphi_{3}$ & $\xi$ & $\xi'$ & $\varphi_1$\\ \hline
				$\!SU(5)\!$ & ${\bf 24}$ & ${\bf 1}$ & ${\bf \overline{5}}$ & ${\bf 10}$ & ${\bf 10}$ & ${\bf 10}$ & ${\bf 5}$ & ${\bf \overline{5}}$ & ${\bf \overline{45}}$ & ${\bf 1}$ & ${\bf 1}$ & ${\bf 1}$ & ${\bf 1}$ & ${\bf 1}$ & ${\bf 1}$\\ \hline
				$A_4$ & ${\bf 1}$ & ${\bf 1}$ & ${\bf 3}$ & ${\bf 1}$ & ${\bf 1}$ & ${\bf 1}$ & ${\bf 1}$ & ${\bf 1}$ & ${\bf 1}$ & ${\bf 3}$ & ${\bf 3}$ & ${\bf 3}$ & ${\bf 1}$ & ${\bf 1}$ & ${\bf 3}$ \\ \hline
				$\!{U}(1)_R\!$ & $1$ & $1$ & $1$ & $1$ & $1$ & $1$ & $0$ & $0$ & $0$ & $0$ & $0$ & $0$ & $0$ & $0$ & $0$ \\ \hline\hline
				$U(1)$ & $-1$ & $2$ & $0$ & $4$ & $1$ & $0$ & $0$ & $0$ & $2$ & $1$ & $-2$ & $0$ & $-1$ & $-4$ & $q_1$ \\ \hline
				$Z_2^1$ & $-$ & $-$ & $+$ & $+$ & $+$ & $+$ & $+$ & $-$ & $-$ & $-$ & $-$ & $-$ & $+$ & $-$ & $+$ \\ \hline
                                $Z_2^2$ & $+$ & $+$ & $+$ & $+$ & $+$ & $-$ & $+$ & $+$ & $+$ & $+$ & $+$ & $-$ & $+$ & $+$ & $+$ \\ \hline
		\end{tabular}
	\caption{Matter, Higgs and flavon chiral superfields in the model. The
          $U(1)$ charge $q_1$ can take any value which prevents $\varphi_1$
          from significantly interacting with the other fields of the model,
          for instance $q_1=-\frac{126}{24}$ as discussed below.}
\label{tab:indirect}
\end{table}

The $U(1)_R$ represents an $R$-symmetry. Its $Z_2$ subgroup gives rise to the
standard $R$-parity which forbids unwanted operators contributing to
proton decay and keeps the lightest SUSY particle a good candidate for cold
dark matter. Moreover, $U(1)_R$ is essential in forbidding $F$-term
contributions to the flavon superpotential which otherwise could dominate the
relevant $D$-term operators used for obtaining the desired vacuum alignment
(see below and the discussion in \cite{Howl:2009ds}).
The $U(1)$ and the two $Z_2$ symmetries constrain the structure of the Yukawa
matrices in the quark and charged lepton sectors. The standard MSSM
$\mu$-term\footnote{Where $H_u$ is the SM doublet of $H_{\bf 5}$; and $H_d$ is
  a linear combination of the SM doublets in $H_{{\bf \overline{5}}}$
  and~$H_{{\bf \overline{45}}}$.} $\mu H_u H_d$ is forbidden by the first of
the $Z_2$ symmetries as well as by $U(1)_R$, allowing for a natural solution
to the $\mu$-problem of the MSSM using a GUT singlet from the hidden sector of
Supergravity theories \cite{Giudice:1988mu}.

The flavon fields $\varphi_i$, $\xi$ and $\xi'$ break the $A_4$
symmetry and constrain the form of the lepton and down quark Yukawa
matrices. The vacuum alignments of the triplet flavon VEVs
that we assume in this model are displayed in Table~\ref{tab:flavons}.
They are achieved using the $D$-term vacuum alignment mechanism
discussed recently in \cite{Howl:2009ds}. This mechanism is ideally suited for
models such as this in which the flavons are
used to generate the neutrino flavour symmetry as an indirect result of the
$A_4$ symmetry as discussed in \cite{King:2009luhn2}. Moreover, the $D$-term
vacuum alignment mechanism does not involve the introduction of extra
``driving fields'' in the superpotential and does not impose any restrictions
on the model other than the requirement that higher order terms in the flavon
potential do not spoil the vacuum alignment arising from the $D$-terms. This
has been demonstrated to arise in a fairly generic way in \cite{Howl:2009ds}
providing that the model also respects a $U(1)_R$ symmetry and involves no
superfields with $R=2$ which, like driving fields, could appear linearly in
the superpotential and lead to large terms in the flavon potential. The
present model involves only fields with $R=0,1$ and so the $D$-term flavon
potential will not receive large corrections from the superpotential. Since
the $D$-term vacuum alignment mechanism is generic and does not provide any
other restrictions on the model than those stated, in this letter we shall simply
assume that this mechanism is in operation, leading to the stated alignments
for $\varphi_{123}, \varphi_{23}, \varphi_3, \varphi_1$.

\begin{table}
	\centering
		\begin{tabular}{|c|c|}
			\hline
				Flavon VEV & VEV alignment\\ \hline
				$\left\langle\varphi_1\right\rangle$ & $(1,0,0)^T$ \\ \hline
				$\left\langle\varphi_3\right\rangle$ & $(0,0,1)^T$ \\ \hline
				$\left\langle\varphi_{23}\right\rangle$ & $\frac{1}{\sqrt{2}}(0,1,-1)^T$ \\ \hline
				$\left\langle\varphi_{123}\right\rangle$ & $\frac{1}{\sqrt{3}}(1,1,1)^T$ \\ \hline
		\end{tabular}
	\caption{The vacuum alignments of the triplet flavons used in
          the model. Without loss of generality, the alignments are given
          without phases; the relative sign between
          $\left\langle\varphi_{23}\right\rangle_2$ and
          $\left\langle\varphi_{23}\right\rangle_3$ is relevant, though the
          actual position of the minus sign is mere convention.}
	\label{tab:flavons}
\end{table}

In order to
avoid the massless Goldstone boson associated with the spontaneously broken
$U(1)$ symmetry, we assume it to be 
gauged.\footnote{If it were not gauged, Goldstone
  boson masses could arise from explicit $U(1)$ breaking in the hidden sector which could generate soft SUSY 
  breaking terms involving only flavon fields where such terms explicitly violate the $U(1)$.
  However such terms could jeopardise the $D$-term alignment mechanism so here we prefer to gauge
  the $U(1)$ to avoid any potential problems.} 
In addition to the particle
content specified in Table~\ref{tab:indirect} we must then introduce extra
matter to cancel the respective gauge anomalies. The cubic $SU(5)$ anomaly
requires the introduction of a Higgs field $H_{\bf 45}$ whose $U(1)$ charge is
determined by the mixed $SU(5)-SU(5)-U(1)$ anomaly to be $q(H_{\bf
45})=-\frac{53}{24}$. Finally the cubic $U(1)$ anomaly can be removed in
many ways; for example, choosing $q_1=-\frac{126}{24}$ we can add three extra
$A_4\times SU(5)$ singlets with $U(1)$ charges $\frac{5}{24}$, $\frac{25}{24}$,
$\frac{51}{24}$. Assuming that $H_{\bf {45}}$ has the same $Z_2$ charges as
$H_{\bf \overline{45}}$ while the three extra $A_4\times SU(5)$ singlets are
neutral under both $Z_2$ symmetries, we have checked that these additional
fields lead to only negligible contributions to the fermion mass matrices
discussed below, provided they get VEVs of order $\epsilon \Lambda$ or
smaller, see Eq.~\eqref{eqn:epsilon}.

\label{sec:model}

\subsection{Allowed terms}

The neutrino sector is composed of Dirac and Majorana mass terms which take
the form in the superpotential:
\begin{equation}
{W}_{\nu}\!=\!\frac{\varphi_{123}}{\Lambda}cF{\bf \psi_{24}}H_{\bf 5}\!+\!\frac{\varphi_{23}}{\Lambda}pFNH_{\bf 5}
\!+\!\frac{\varphi_{23}^2}{2\Lambda}y_N NN\!+\!\frac{\xi^4}{2\Lambda^3}y'_N NN\!+\!\frac{\varphi_{123}^2}{2\Lambda}y \mathrm{Tr}\left({\bf \psi_{24}}^2\right),
\label{eqn:neutrino}
\end{equation}
with $\Lambda$ a heavy mass scale and $c,p,y_N,y'_N,y$ dimensionless coupling
constants.  When the flavons get their VEVs the superpotential in
Eq.~\eqref{eqn:neutrino} reproduces that in Eq.~\eqref{eqn:type3} but with
constrained couplings $c_i$ and $p_i$ leading to TB mixing.

The superpotential terms of the down quark and charged lepton sector are given
as follows
\begin{equation}
{W}_d\sim\frac{\varphi_{23}\xi^2}{\Lambda_d^3}T_1FH_{{\bf \overline{5}}}+\frac{\varphi_{123}\xi^2}{\Lambda_d^3}T_2F H_{{\bf\overline{5}}}+\frac{\varphi_{23}\xi}{\Lambda_d^2}T_2FH_{{\bf \overline{45}}}+\frac{\varphi_{3}}{\Lambda_d}T_3FH_{{\bf \overline{5}}},
\label{eqn:down}
\end{equation}
where $\Lambda_d$ is the relevant messenger mass. The flavon $\xi$ plays a role
similar to a Froggatt-Nielsen field \cite{Froggatt:1979heirarchy}, except that
it is not the sole contributor to the generated mass hierarchy, here combined
as it is with the triplet flavons.

Finally the up quark sector Yukawa superpotential terms take the form
\begin{equation}
\begin{split}
{W}_u&\sim\frac{(\xi')^2}{\Lambda_u^2}T_1T_1H_{\bf 5}+\left(\frac{\varphi_{23}^2\xi}{\Lambda_u^3}+\frac{\xi^5}{\Lambda_u^5}\right)(T_1T_2+T_2T_1)H_{\bf 5}+\frac{\varphi_{23}\varphi_{3}\xi^2}{\Lambda_u^4}(T_1T_3+T_3T_1)H_{\bf 5} \\
&+\frac{\xi^2}{\Lambda_u^2}T_2T_2H_{\bf 5}+\frac{\varphi_{123}\varphi_3\xi^2}{\Lambda_u^4}(T_2T_3+T_3T_2)H_{\bf 5}+T_3T_3H_{\bf 5}.
\end{split}
\label{eqn:up}
\end{equation}
It should be mentioned that the messenger mass in this sector, $\Lambda_u$,
may in principle be different from that in the down quark sector. The field
$\xi'$ is introduced specifically to generate the $T_1T_1$ term to the
required order.

\subsection{Fermion mass matrices}
\label{sec:masses}
After spontaneous breakdown of the $A_4$ family symmetry by the flavon VEVs,
the superpotential terms of Eqs.~\eqref{eqn:neutrino}, \eqref{eqn:down} and
\eqref{eqn:up} predict mass matrices for the respective sectors. In the
following, order one coefficients in the quark and charged lepton sectors are
omitted (including flavon VEV normalisation factors). Regarding the scale of
the flavon VEVs we define
\begin{equation}
\eta_i = \frac{\left\langle|\varphi_i|\right\rangle}{\Lambda} ,
\label{eqn:eta}
\end{equation}
where $\varphi_i=\varphi_{123}$, $\varphi_{23}$, $\varphi_{3}$, $\xi$ or
$\xi'$. In order to get the hierarchical structure of the quark and charged
lepton mass matrices we assume\footnote{It is possible to have a hierarchy in
  the flavon VEVs since the scales at which their mass terms are driven
  negative can vary \cite{Howl:2009ds}.}
\begin{equation}
\eta_{123},\eta_{23},\eta_{\xi'}=\epsilon^2\;
~~\mathrm{and}\;~~ \eta_3,\eta_{\xi}=\epsilon ,
\label{eqn:epsilon}
\end{equation}
where the numerical values for $\epsilon$ depend on the messenger scale of the
relevant sector. We note that we have given the superpotential terms of the
quark and charged lepton sectors up to and including $\mathcal O(\epsilon^5)$.

In the Higgs sector, it is not the $H_{\bf 5}$, $H_{{\bf \overline{5}}}$ or
$H_{{\bf \overline{45}}}$ which get VEVs but their SM doublet
components. These are the two MSSM doublets $H_u$ (corresponding to
$H_{\bf 5}$) and $H_d$ (corresponding to a linear combination of
$H_{{\bf \overline{5}}}$ and $H_{{\bf \overline{45}}}$); they originate below the
GUT scale and remain massless down to the electroweak scale. The non-MSSM
states all acquire GUT scale masses, including the linear combination of
$H_{{\bf \overline{5}}}$ and $H_{{\bf \overline{45}}}$ orthogonal to
$H_d$. Electroweak symmetry is broken after the light MSSM doublets $H_{u,d}$
acquire VEVs $v_{u,d}$ and they then generate the fermion masses.

In the following all quark and charged lepton mass matrices are given in the
L-R convention, i.e. the mass term for a field $\psi$ is given in the order
$\psi_L  M_{LR}\psi_R$.

\subsubsection{Neutrino sector}
In our model the light neutrino masses arise from a combination of type~I and
type~III seesaw. Due to the absence of a $H_{\bf 24}$ the heavy seesaw messenger
particles $N$ and $\rho_0$ do not mix as can be seen from
Eq.~(\ref{eqn:neutrino}). Thus the $2\times 2$ Majorana mass matrix of the
heavy right-handed  $SU(2)_L$ singlets is automatically diagonal. Furthermore, 
the seesaw messenger responsible for the type~III contribution, $\rho_3$, 
cannot mix with $N$ as they furnish different $SU(2)_L$ representations.
In CSD the (approximate) diagonal nature of the seesaw particles is usually
a necessary extra assumption which often lacks a fundamental explanation. In
our adjoint model, however, it is directly built into the theory by not
including  $H_{\bf 24}$. Therefore our model represents a very natural
realisation of CSD.

In the Dirac neutrino sector of Eq.~\eqref{eqn:neutrino}, the spontaneous
breaking of the $A_4$ family symmetry by the flavon VEVs $\langle
\varphi_{123}\rangle$ and $\langle \varphi_{23}\rangle$ gives
\begin{equation} \mathscr{L}_{\nu}=\frac{c\eta_{123}v_u}{\sqrt{3}}
(\nu_e+\nu_{\mu}+\nu_{\tau})
\left(\frac{\rho^{0}_3}{2}-\sqrt\frac{3}{20}\,\rho_0\right)
-\frac{p\eta_{23}v_u}{\sqrt{2}}(\nu_{\mu}-\nu_{\tau})N
~+~\text{h.c.}\
\label{eqn:typeIIIbexpanded} ,
\end{equation}
where the numerical factors of $\rho^0_3$ and $\rho_0$ are determined from the
normalised $SU(5)$ generators in the adjoint representation
\cite{Langacker:1980js}. Upon application of the seesaw formula of
Eq.~\eqref{eqn:type3matrix} we find the effective left-handed Majorana
neutrino mass matrix
\begin{equation}
\label{eqn:numasses}
M_{\nu}=\frac{2c^2v_u^2}{15y\Lambda}\begin{pmatrix}
																																						 1 & 1 & 1 \\
																																						 1 & 1 & 1 \\
																																						 1 & 1 & 1
																																						 \end{pmatrix}
			 +\frac{p^2v_u^2}{2(y_N+y'_N\eta_{\xi}^4/\eta_{23}^2)\Lambda}\begin{pmatrix}
			 															0 & 0 & 0 \\
			 															0 & 1 & -1 \\
			 															0 & -1 & 1
			 															 \end{pmatrix}.
\end{equation}
Since any matrix diagonalisable by Eq.~\eqref{eqn:tbm} may be written
as\footnote{$\varphi'_{1}\propto \frac{1}{\sqrt{6}}(-2,1,1)^T$.}
$m_1\varphi'_{1}\left(\varphi'_{1}\right)^T/|\varphi_1'|^2+m_2\varphi_{123}(\varphi_{123})^T/|\varphi_{123}|^2+m_3\varphi_{23}(\varphi_{23})^T/|\varphi_{23}|^2$ \cite{King:2009luhn2}, we may readily read off
the masses and state that
\begin{equation}
\label{eqn:masses}
M_{\nu}^{\mathrm{diag}}=\begin{pmatrix}
				0 & 0 & 0 \\
				0 & m_2 & 0 \\
				0 & 0 & m_3
		            \end{pmatrix},
\;\;\;\mathrm{with}\;\;\;m_2=\frac{2c^2v_u^2}{5y\Lambda}\ ,\quad m_3=\frac{p^2v_u^2}{(y_N+y'_N{\eta_{\xi}^4}/{\eta_{23}^2})\Lambda}.
\end{equation}
Hence the model predicts one massless left-handed neutrino and thus a
hierarchical neutrino mass spectrum.

\subsubsection{Down quark and charged lepton sector}
In the down quark and charged lepton sector, the superpotential of
Eq.~\eqref{eqn:down} predicts a mass matrix of the form (with messenger mass
$\Lambda_d$ in $\eta_i$)
\begin{equation}
 \begin{pmatrix}
      0 & \eta_{23}\eta_{\xi}^2 & -\eta_{23}\eta_{\xi}^2 \\
      \eta_{123}\eta_{\xi}^2 & \eta_{123}\eta_{\xi}^2 +k_f\eta_{23}\eta_{\xi} & \eta_{123}\eta_{\xi}^2-k_f\eta_{23}\eta_{\xi} \\
      0 & 0 & \eta_3
     \end{pmatrix}v_d,
\label{eqn:downmass1}
\end{equation}
where $k_f$ is the Georgi-Jarlskog factor (in the case that $f=e$, the mass
matrix must also be transposed):
\begin{displaymath}
 k_f=\begin{cases}
      1 &\text{for} \quad f=d, \\
      -3 &\text{for} \quad f=e.
     \end{cases}
\end{displaymath}
Inserting the $\epsilon$ suppressions of the flavon VEVs from
Eq.~(\ref{eqn:epsilon})  the down quark mass matrix becomes
\begin{equation}
 M_d\sim\begin{pmatrix}
         0 & \epsilon^3 & -\epsilon^3 \\
	 \epsilon^3 & \epsilon^2 & -\epsilon^2 \\
	 0 & 0 & 1
        \end{pmatrix}\epsilon \,v_d
\label{eqn:downmass2},
\end{equation}
whilst the charged lepton mass matrix reads
\begin{equation}
 M_e\sim\begin{pmatrix}
         0 & \epsilon^3 & 0 \\
	 \epsilon^3 & -3\epsilon^2 & 0 \\
	 -\epsilon^3 & 3\epsilon^2 & 1
        \end{pmatrix}\epsilon\, v_d
\label{eqn:chargedmass}.
\end{equation}
Here we assume the numerical value $\epsilon\sim0.15$.
Upon diagonalisation, these give mass ratios of $\epsilon^4:\epsilon^2:1$ for
the down quarks and $\frac{\epsilon^4}{3}:3\epsilon^2:1$ for the charged
leptons. These ratios are in good agreement with quark and lepton data and
also predict GUT scale mass relations of $m_e \sim \frac{m_d}{3}$,
$m_{\mu}\sim 3m_{s}$ and $m_{\tau}\sim m_b$ as desired. In the low quark angle
approximation, left-handed down quark mixing angles
$\theta^d_{12}\sim\epsilon$, $\theta^d_{13}\sim\epsilon^3$ and
$\theta^d_{23}\sim\epsilon^2$ are also predicted in agreement with data
(assuming an approximately diagonal up sector which we obtain in the next
Section). The corresponding charged lepton mixing angles are
$\theta^e_{12}\sim\frac{\epsilon}{3}$, $\theta^e_{13}\sim0$ and
$\theta^e_{23}\sim0$.

The Pontecorvo-Maki-Nakagawa-Sakata (PMNS) matrix is not of exact
TB form but receives small corrections from charged lepton
mixing. In particular, the reactor angle deviates from zero by $\theta_{13}
\sim \frac{1}{\sqrt{2}} \frac{\epsilon}{3}$ \cite{sumrule}. 
Furthermore, since $\theta^e_{13}\sim \theta^e_{23}\sim 0$,
two sum rules for lepton mixing are
respected~\cite{sumrule,Antusch:2008yc}. Expressed in terms of the
(r)eactor, (s)olar and (a)tmospheric deviation parameters 
defined as $\sin \theta_{13} = \frac{r}{\sqrt{2}}$, $\sin \theta_{12} =
\frac{1}{\sqrt{3}}(1+s)$, $\sin \theta_{23} = \frac{1}{\sqrt{2}}(1+a)$
\cite{King:2007pr}, the sum rules read $s=r\cos \delta$ and $a=-r^2/4$
\cite{S4-LQsum}, with $\delta$ being the leptonic Dirac CP phase. 

\subsubsection{Up quark sector}
Eq.~\eqref{eqn:up} may be expanded after $A_4$
symmetry breaking and is responsible for up quark masses:
\begin{equation}
 \begin{pmatrix}
      \eta_{\xi'}^2 & \eta_{23}^2\eta_{\xi}+\eta_\xi^5 & -\eta_{23}\eta_3\eta_{\xi}^2 \\
      \eta_{23}^2\eta_{\xi}+\eta_\xi^5 & \eta_{\xi}^2 & \eta_{123}\eta_3\eta_{\xi}^2 \\
      -\eta_{23}\eta_3\eta_{\xi}^2 & \eta_{123}\eta_3\eta_{\xi}^2 & 1
     \end{pmatrix}v_u.
\label{eqn:upmass1}
\end{equation}
Taking the VEV hierarchy as in Eq.~\eqref{eqn:epsilon}, but now adopting the
messenger scale $\Lambda_u \approx 3 \Lambda_d$, we obtain a mass matrix with
an expansion parameter $\overline{\epsilon}\sim0.05$,
\begin{equation}
 M_u\sim\begin{pmatrix}
      \overline{\epsilon}^4 & \overline{\epsilon}^5 & -\overline{\epsilon}^5 \\
      \overline{\epsilon}^5 & \overline{\epsilon}^2 & \overline{\epsilon}^5 \\
      -\overline{\epsilon}^5 & \overline{\epsilon}^5 & 1
     \end{pmatrix}v_u.
\label{eqn:upmass2}
\end{equation}
and an up quark mass hierarchy
$\overline{\epsilon}^4:\overline{\epsilon}^2:1$. As the mass matrix of
Eq.~(\ref{eqn:upmass2}) is diagonal to a good approximation, the up quark
mixing is negligible. An important consequence of this observation is that the
Cabibbo-Kobayashi-Maskawa (CKM) mixing arises predominantly from the down
quark sector, with the Cabibbo angle being $\theta_C \sim \theta_{12}^d \sim
\epsilon$.

\section{Conclusions}
\label{sec:conclusion}

In conclusion, minimal (SUSY) $SU(5)$ represents an attractive route to
unification, but the Weinberg operator cannot account for neutrino mass and
mixing, and the seesaw mechanisms all require extra matter or Higgs below the
GUT scale. An appealing possibility, considered here, is to extend SUSY $SU(5)$
by assuming a single right-handed neutrino singlet and an adjoint matter
representation below the GUT scale, 
including an $A_4$ family symmetry as well as a gauged anomaly-free $U(1)$.
Hierarchical neutrino masses result from a combined type~I and type~III
seesaw mechanism, and TB mixing arises indirectly from the $A_4$ family
symmetry. 

One attractive feature of this scheme is that the mixing between the
single right-handed neutrino and the matter in the adjoint
can be forbidden by not including the $H_{\bf 24}$, leading to a diagonal
heavy Majorana sector as required by CSD.
The flavon vacuum alignments arise from the elegant SUSY $D$-term mechanism.
The model also reproduces
a realistic description of quark and charged lepton masses and quark mixings,
including the Georgi-Jarlskog relations.

Corrections to TB mixing in the lepton sector come solely from the
1-2~mixing of the left-handed charged leptons, resulting in a PMNS matrix
which is within the experimentally allowed limits. In particular the model
respects the sum rules $s=r\cos \delta$ and $a=-r^2/4$ with $r=\theta_C/3$.

\section*{Acknowledgments}
We thank Pavel Fileviez Perez for discussions in the initial stage of this
work. SFK and CL acknowledge support from the STFC Rolling Grant ST/G000557/1.
SFK is grateful to the Royal Society for a Leverhulme Trust Senior Research
Fellowship and a Travel Grant.




\providecommand{\bysame}{\leavevmode\hbox to3em{\hrulefill}\thinspace}

\end{document}